\documentclass[12pt]{article}
\usepackage{latexsym}
\usepackage{amsmath,,calrsfs}
\usepackage{amsfonts}
\usepackage{amssymb}
\usepackage{amscd}
\usepackage{bbm}
\usepackage{fancybox}
\usepackage{cite}
\usepackage{amsmath,amsfonts,amsbsy}
\usepackage{pstricks,pst-node}
\usepackage[small,bf,hang]{caption2}
\usepackage{graphicx}
\usepackage{epsfig}
\usepackage{psfrag}
\usepackage{comment}
\usepackage{color}

\usepackage{float}

\psset{unit=1.3cm,linewidth=.5pt,radius=.2}  

\usepackage{multirow}                     
\usepackage{float}                          
\usepackage{lscape}                         
\usepackage{bm}

\newcommand{\bb}{\bar\beta}

\newcommand{\beq}{\begin{equation}}
\newcommand{\eeq}{\end{equation}}
\newcommand{\bi}{\begin{itemize}}
\newcommand{\ei}{\end{itemize}}
\newcommand{\bt}{\begin{tabular}}
\newcommand{\et}{\end{tabular}}
\newcommand{\bc}{\begin{center}}
\newcommand{\ec}{\end{center}}

\newcommand{\be}{\begin{equation}}
\newcommand{\ee}{\end{equation}}
\newcommand{\bea}{\begin{eqnarray}}
\newcommand{\eea}{\end{eqnarray}}
\newcommand{\ba}{\begin{array}}
\newcommand{\ea}{\end{array}}

\def\bbox{{\,\lower0.9pt\vbox{\hrule \hbox{\vrule height 0.2 cm
\hskip 0.2 cm \vrule height 0.2 cm}\hrule}\,}}
\newcommand{\dsl}{\pa \kern-0.5em /}

\font\mybb=msbm10 at 12pt
\def\bb#1{\hbox{\mybb#1}}

\def\bR {\bb{R}}

\def\bT {\bb{T}}

\def\bC {\bb{C}}

\def\bI{\bb{I}}

\def\bfG{\mbox{\boldmath $\Gamma$}}

\def\bfpi{\mbox{\boldmath $\pi$}}

\def\bfG{\mbox{\boldmath $\Gamma$}}
\def\bfg{\mbox{\boldmath $\gamma$}}

\def\bfpi{\mbox{\boldmath $\pi$}}

\makeatletter \@addtoreset{equation}{section} \makeatother

\def\slashchar#1{\setbox0=\hbox{$#1$}           
   \dimen0=\wd0                                 
   \setbox1=\hbox{/} \dimen1=\wd1               
   \ifdim\dimen0>\dimen1                        
      \rlap{\hbox to \dimen0{\hfil/\hfil}}      
      #1                                        
   \else                                        
      \rlap{\hbox to \dimen1{\hfil$#1$\hfil}}   
      /                                         
   \fi}

\begin{document}

\begin{titlepage}
\begin{center}

\hfill  DAMTP-2016-81, ICCUB-16-039, UTTG-24-16

\vskip 1.5cm

{\Large \bf  The Galilean Superstring}

\vskip 2cm

{\bf Joaquim Gomis${}^{1,}{}^2$  and  Paul K.~Townsend\,${}^3$} \\

\vskip 5pt

{\em $^1$ \hskip -.1truecm
\em  Departament de F{\i}sica Qu\`antica i Astrof{\i}sica\\ and Institut de Ci\`encies del Cosmos (ICCUB), 
Universitat de Barcelona, \\Mart\i i Franqu\`es 1, E-08028 Barcelona, Spain  \vskip 5pt }

{\em $^2$ \hskip -.1truecm
\em  Theory Group, Department of Physics, University of Texas\\ 
Austin, TX, 78712 \vskip 5pt }

{email: {\tt gomis@ecm.ub.edu}} \\

\vskip .4truecm

{\em $^3$ \hskip -.1truecm
\em  Department of Applied Mathematics and Theoretical Physics,\ Centre for Mathematical Sciences, University of Cambridge,\\
Wilberforce Road, Cambridge, CB3 0WA, U.K.\vskip 5pt }

{email: {\tt P.K.Townsend@damtp.cam.ac.uk}} \\

\end{center}

\vskip 0.5cm

\begin{center} {\bf ABSTRACT}\\[3ex]
\end{center}

The action for a Galilean superstring is found from  a non-relativistic  limit of the closed Green-Schwarz (GS) superstring; it has zero tension and
provides an example of a massless super-Galilean system   A Wess-Zumino term leads to a topological  central charge in the Galilean 
supersymmetry algebra, such that unitarity requires a upper bound on the total momentum. This Galilean-invariant bound, which is also implied by the 
classical phase-space constraints,  is saturated by solutions of the superstring equations of motion that half-preserve supersymmetry. 
We discuss briefly the extension to the Galilean supermembrane.

\end{titlepage}

\newpage
\setcounter{page}{1} 
\tableofcontents


\section{Introduction}

The non-relativistic limit of relativistic particle mechanics is a limit in which all velocities become negligible when compared to $c$, the speed of light.
Formally, this can be viewed as a  limit in which $c\to\infty$ but in other contexts  there may be several such limits, according to what is kept fixed as 
$c\to\infty$. One  context in which different non-relativistic limits are possible is the mechanics of extended objects. In this paper we focus on the 
non-relativistic limit of a closed relativistic string.

One non-relativistic limit was described in \cite{Gomis:2000bd, Danielsson:2000gi,Gomis:2004pw,Brugues:2004an,Gomis:2005pg};  it is a limit in which a small piece of 
the string worldsheet becomes a $(1+1)$-dimensional Minkowski vacuum for  small fluctuations of the string. This is a special case of a limit applying  
to a relativistic $p$-brane  that reduces to  the usual low-velocity limit for $p=0$, whereas only transverse velocities are assumed small for $p>0$.
This was called the ``field-theory'' limit in \cite{Townsend:1999hi} since small fluctuations of the Minkowski ``brane vacuum''  still travel at the speed of 
light within the brane.

Recently, a different non-relativistic limit  was described \cite{Batlle:2016iel}.  It applies for any $p>0$  but here we restrict to $p=1$ except for some concluding remarks.
The starting point is the Nambu-Goto (NG) string of tension $T$. For simplicity, we assume here that the string is closed and of unit parameter length.  For 
Minkowski coordinates $X = (ct,{\bf x})$, and 
including all factors of  $c$,  the action is 
\begin{equation}
S_{NG} =- (T/c)\int\! d\tau\! \oint \! d\sigma  \, \sqrt{ (\dot X\cdot X^\prime)^2- \dot X^2 X^\prime{}^2} \, . 
\end{equation}
As usual, an overdot indicates a partial derivative with respect to worldsheet time $\tau$ and a prime indicates a partial derivative with respect to the string coordinate  
$\sigma\sim \sigma + 1$.  Taking the $c\to\infty$ limit yields what we shall call here the ``Galilean string''  action:
\begin{equation}\label{NRS}
S= -T\int \! d\tau\! \oint \! d\sigma \,  \sqrt{|\dot t{\bf x}' - t'\dot{\bf x}|^2}\, .
\end{equation}

As $c$ has dimensions of velocity, the $c\to\infty$ limit should be viewed as one for which some other velocity $v$ is held fixed, in which case the limit is one for which $v/c \to 0$, but what 
is this other velocity $v$? We answer this question below, but we mention here that there is an alternative way to understand the Galilean limit, 
which was adopted in \cite{Batlle:2016iel}. One first rescales $t\to\omega t$, where $\omega$ is dimensionless.   After this rescaling,  the dependence of the NG action on  $c$ and $T$ is replaced by a dependence on $\omega c$ and $\omega T= \tilde T$, so the limit $\omega\to\infty$ for fixed $\tilde T$ yields the above Galilean string action but with  $\tilde T$ replacing $T$.

 The phase-space form of the Galilean string action is  \cite{Batlle:2016iel}
\begin{equation}\label{psact1}
S= \int\! d\tau\! \!\oint \! d\sigma \, \left\{ \dot{\bf x}\cdot{\bf p} - \dot t E - \mu\left[ {\bf x}'\cdot{\bf p} - t'E\right] - \frac{1}{2} \lambda\left[|{\bf p}|^2 - (T t')^2\right]\right\}\, ,
\end{equation}
where $\mu$ and $\lambda$ are two Lagrange multipliers for the two phase-space constraints associated with invariance under worldsheet reparametrizations.  
The equivalence with (\ref{NRS}) may be verified by first eliminating 
${\bf p}$ by its equation of motion and then eliminating $(E,\mu)$ by the  $(\mu,E)$ equations of motion. This leads to an action with Lagrangian density
$\tfrac{1}{2}[|\dot t {\bf x}'-t'\dot{\bf x}|^2/(\lambda t'{}^2) + \lambda (Tt')^2]$.  Varying with respect to $\lambda$ now yields two solutions for $\lambda$, which differ by a sign, and back-substitution yields (\ref{NRS}) if we allow $T$ to have either sign.  

The phase-space action (\ref{psact1}) is Galilean invariant with Noether charges
\begin{eqnarray}\label{Ns}
H = \oint \! d\sigma \, E\, , \quad {\bf P} = \oint \! d\sigma \, {\bf p}\, \, , \quad {\bf B} = \oint\! d\sigma \, t{\bf p}\, , \quad
{\bf J} = \oint \! d\sigma\, {\bf x}\times {\bf p} \, . 
\end{eqnarray}
The Poisson bracket algebra of these charges is the Galilean algebra. A comparison with the algebra of Galilean charges for the standard 
non-relativistic point particle is instructive: in that case the particle's mass appears as a central charge in the Poisson bracket of ${\bf P}$ with ${\bf B}$,  thereby enlarging the  Galilei algebra to the Bargmann algebra. There is no such central charge in the Galilean string algebra, so the Galilean string is an example of a ``massless Galilean  system'' \cite{Sou,GS,Duval:2009vt}. 
 
The Galilean algebra spanned by the Noether charges (\ref{Ns}) is a subalgebra of the algebra spanned by those functions $f(t,E; {\bf x},{\bf p})$ for which 
\begin{equation}
\left\{f, \varphi\right\}_{PB}  \propto \varphi \, , \qquad \varphi := |{\bf p}|^2 - (T t')^2\, . 
\end{equation}
In other words, the full symmetry algebra is the subalgebra of the algebra of symplectic difeomorphisms that preserves the constraint $\varphi=0$ that is imposed by the Lagrange multiplier
$\lambda$.  This is an infinite dimensional algebra that contains the (still infinite-dimensional) subalgebra of ``point-transformations'' studied in \cite{Batlle:2016iel}. 
 
 The phase-space action of the Galilean string makes it apparent that the worldline-time reparametrization invariance cannot be
fixed by the gauge choice $t(\tau,\sigma)\propto \tau$, as would have been possible prior to the Galilean limit.  However, we can fix the 
 $\sigma$-reparametrization invariance by the gauge choice $t\propto \sigma$, although the global validity of this choice for a closed string requires a periodic identification of $t$.  
We shall assume this and choose time units such that 
\begin{equation}\label{idt}
t \sim  t+1\, .  
\end{equation}
This does not mean that our starting point should have been the NG string in a Minkowki space subject to periodic identification of the Minkowski
time coordinate $x^0=ct$ because a finite period  for $t$ becomes an infinite period for $x^0$ when $c\to\infty$. Rather, the possibility of a periodic
identification of  $t$, {\it without violation of boost invariance}, arises after the $c\to \infty$ limit has been taken; this is a feature of the Galilean string 
that is absent from the NG string.  Once this identification is made, we have the integer topological charge 
\begin{equation}
n= \oint \! d\sigma \, t' \, , 
\end{equation}
which is the winding number of the map from the string to the $t$-circle. For the choice $n=1$, which we will assume for most of what follows, 
we may choose the gauge 
\begin{equation}
t(\tau,\sigma) = \sigma\, . 
\end{equation}
We may now solve the constraint imposed by $\mu$ for $E$,  with the result that $E= {\bf x}'\cdot{\bf p}$. This formula is relevant to
Noether charge $H$ but it is not needed for the action, which becomes
\begin{equation}\label{wind1}
S= \int\! d\tau\! \!\oint \! d\sigma \, \left\{ \dot{\bf x}\cdot{\bf p} -  \frac{1}{2} \lambda\left(|{\bf p}|^2 - (nT)^2\right)\right\}\, .
\end{equation}
The remaining constraint, due to time-reparametrization invariance,  imposes an upper bound on the magnitude of the total $3$-momentum ${\bf P}$. 
In fact, this bound is already a consequence of the time-reparametrization constraint prior to any gauge fixing:  integration of it yields
\begin{equation}
T^2 \oint \! d\sigma\, t'{}^2  = \oint \! d\sigma |{\bf p}|^2 \ge |{\bf P}|^2\, , 
\end{equation}
where the inequality, which holds for any choice of $t(\sigma)$, follows from positivity of $\oint\! d\sigma |{\bf p -P}|^2$.  
We get the strongest bound, for given string winding number  $n$, by minimizing $\oint \! d\sigma\, t'{}^2$, and positivity of $\oint\! d\sigma (t' -n)^2$ implies that this minimum is $n^2$, 
realized when $t'=n$.  For $n=1$ this yields the bound
\begin{equation}\label{bogbound}
|{\bf P}|^2 \le T^2\,  \qquad (n=1). 
\end{equation}
This bound  is saturated when ${\bf p}(\sigma)={\bf P}$ for a string parametrization such that $t'=1$.  The bound is boost invariant because, as mentioned above, 
${\bf P}$ has zero Poisson bracket with the boost generator ${\bf B}$.

The action (\ref{wind1}) is still invariant under reparametrizations of worldsheet time. At least locally,  this allows us to identify  one of the space 
coordinates with $\tau$, and we may then choose the unit of length such that 
\begin{equation}\label{gfix}
{\bf n}\cdot {\bf x} =  \tau\, ,  
\end{equation}
for some unit space vector ${\bf n}$.  This gauge choice is legitimate provided ${\bf n}\cdot{\bf p} \ne 0$, and given this we may solve the 
remaining constraint for  ${\bf n}\cdot{\bf p}$. This yields the physical phase-space action 
\begin{equation}\label{ppsact}
S= \int\! d\tau \!\oint \! d\sigma \, \left\{ \dot{\bf x}_\perp\cdot{\bf p}_\perp - T\sqrt{1- T^{-2} |{\bf p}_\perp|^2} \right\}\, ,  
\end{equation}
where ${\bf x}_\perp$ are the coordinates of the ``tranvserse'' space; i.e. ${\bf n}\cdot {\bf x}_\perp\equiv 0$.  
Although this action does not involve $\sigma$-derivatives,  the Noether charge $H$ does; in terms of the physical phase-space variables it is 
\begin{equation}
H= \oint \! d\sigma\,  {\bf x}'_\perp \cdot {\bf p}_\perp\, . 
\end{equation}
Although $H$ plays the role of energy with respect to Galilean boosts, it is not bounded from either  below or above, and although  $-{\bf n}\cdot {\bf P}$ is the energy 
for the mechanical system with action (\ref{ppsact}), it is not invariant under the full rotation group. The concept of ``energy''  is therefore problematic for the Galilean string.

Because the Hamiltonian density of the action (\ref{ppsact}) is independent of ${\bf x}_\perp$, each string element moves independently of adjacent string elements. 
In this respect, the Galilean string is similar to the tensionless limit of the NG string, hence the ``non-vibrating'' terminology of  \cite{Batlle:2016iel}. 
In fact, as mentioned above, the Galilean string {\it is} tensionless. Another way to see this is by comparison of its physical-gauge equation of motion $\ddot{\bf x}_\perp =0$ 
with the equation for small-amplitude transverse waves on an ideal string of tension $\bT$ and (non-kinetic) energy density ${\cal E}$:
\begin{equation}
\ddot{\bf x}_\perp =  v^2{\bf x}'_\perp{}^\prime \, , \qquad \left(v/c\right)^2 = \bT/{\cal E}\, . 
\end{equation} 
Causality requires $v\le c$, which is  saturated by a NG string because in this case $\bT={\cal E}=T$.  Although $T$ is the tension of the NG string, it could equally well be called the 
energy density,  and that is how it should be interpreted in the $c\to\infty$ limit that leads to the Galilean string. In this limit, $T={\cal E}$ is held fixed, so $\bT\to 0$ is equivalent to a 
$v/c\to 0$ limit where $v$ is the velocity of transverse waves on the string (for the rescaling limit used in \cite{Batlle:2016iel} the rescaled
tension $\tilde T$ should be interpreted as the energy density ${\cal E}$).  This zero tension limit is very different from the limit that  yields the {\it relativistic} tensionless string; 
in that case the tension remains equal to the energy density and both go to zero such that each string element moves independently at the speed of light, which remains finite. 

Another unusual  feature of the Galilean string is that the Hamiltonian density of the physical phase-space action (\ref{ppsact})  is bounded not only from below (by zero for positive $T$) but also {\it from above}.   Moreover,  transverse momentum contributes  {\it negatively} to it (for positive $T$)  such  that the upper bound  is saturated when the transverse momentum density is zero. We could arrange for transverse momentum to contribute positively by choosing $T$ to be negative; the Hamiltonian density  is then negative but  bounded {\it from below}, 
such that it becomes non-negative if we add to it the constant $|T|$. It may be that it makes more physical sense to suppose that $T$ is negative,  but the sign of $T$ will not be of 
importance to the results of this paper, so we pass over this issue. The reader may assume, unless stated otherwise, that $T$ is positive. 

Our main goal here is to present the extension of the Galilean string to the Galilean superstring, which is a non-relativistic limit of the manifestly Lorentz-invariant Green-Schwarz (GS) superstring \cite{Green:1983wt}.  We recall that  the GS action is the sum of two terms. One is a straightforward supersymmetric extension of the NG string, a ``super-NG'' term. The other can be interpreted as a super-Galilean invariant Wess-Zumino term for the supertranslation subgroup \cite{Henneaux:1984mh}, and the relative coefficient is such that the action has a fermionic gauge invariance (``$\kappa$-symmetry'') allowing half of the fermionic variables to be ``gauged away''; i.e. only half of the  spacetime-spinor worldsheet fields are physical \cite{Green:1983wt}. Here we show that  the $c\to\infty$ limit yields a Galilean superstring with similar features. This is also true of the ``non-relativistic superstring'' described in \cite{Gomis:2004pw,Gomis:2005pg}, but the points of similarity are significantly different from the Galilean superstring. 

The analog of the ``super-NG'' term in the GS action is  a super-Galilei invariant extension of the Galilei string action (\ref{NRS}), where the super-Galilei algebra is such that a commutator of two supersymmetry transformations is a space translation. The WZ term of the GS action becomes a WZ term for the super-translation subgroup of the super-Galilei group, and  it leads to a modification of the  supersymmetry algebra by a topological  charge, which is  the product of $T$ with the winding number $n$. This is analogous to the situation for the GS superstring \cite{deAzcarraga:1989mza}, but simpler because the  topological charge of the super-Galilei algebra  is a central charge. This central charge is crucial to unitarity of the quantum theory because it leads to the conclusion that the super-Galilei algebra of quantum Noether charges is compatible with the absence of negative-norm states as a consequence of the classical bound (\ref{bogbound}). 
Moreover, bosonic solutions of the Galilean superstring equations that saturate this bound preserve half of the Galilean supersymmetry.  

We present both general argument for these results, and a verification of them  by a direct computation of the Poisson brackets of the Galilean supersymmetry Noether charges in a physical gauge; the action in this gauge is a Galilean supersymmetric extension of the action (\ref{ppsact}). We also describe the special  features of supersymmetric solutions of the Galilean superstring equations of motion in the context of a discussion of generic solutions. 

We summarize our results in the conclusions and briefly discuss some aspects of their extension to a Galilean super-p-brane for $p>1$.

\section{Galilean superstring}

Our  starting point is now the GS superstring, which exists for spacetime dimensions $D=3,4,6,10$ and in versions with ${\cal N}=1$ and ${\cal N}=2$ spacetime supersymmetry. 
Here we shall choose $D=4$ and ${\cal N}=1$, and leave a brief discussion of other cases to the conclusions. 
For our case, the GS superstring  involves an additional worldsheet field $\theta$ that is an {\it anticommuting} four-component spacetime Majorana spinor.  We will assume
that the Dirac matrices $(\Gamma^0,\bfG)$ are real, in which case Majorana spinors are real and the Dirac conjugate of $\theta$ equals its Majorana  conjugate, which 
is
\begin{equation}
\bar\theta = \theta^T\Gamma^0\, . 
\end{equation} 
The Poincar\'e invariant NG string action is converted into a super-Poincar\'e invariant action by the replacement of the one-forms $cdt$ and $d{\bf x}$ by, respectively, 
\begin{equation}
\pi^0 = cdt  + i \bar\theta \Gamma^0 \theta\, , \qquad \bfpi = d{\bf x} +  i \bar\theta \bfG \theta\, . 
\end{equation}
In the $c\to\infty$ limit this yields the Lagrangian density
\begin{equation}
{\cal L}_{sNG} =-T \sqrt{ |\dot t \bfpi_\sigma - t' \bfpi_\tau |^2}\, , 
\end{equation}
where $\bfpi_\tau$ and $\bfpi_\sigma$ are the components of the pullback of $\bfpi$ to the worldsheet. 
The WZ term in the GS action is constructed from the super-Poincar\'e invariant superspace 3-form \cite{Henneaux:1984mh}
\begin{equation}
 (T/c) \pi^m i\bar d\theta \Gamma_m d\theta  = -\, d\left[ Tdt \, i\bar\theta\Gamma_0 d\theta\right] + {\cal O}(1/c) \, . 
\end{equation}
In the $c\to\infty$ limit this yields the Lagrangian density 
\begin{equation}
{\cal L}_{WZ} = -T \left(\dot t \, i\bar\theta\Gamma_0 \theta' - t'\, i \bar\theta\Gamma_0 \dot\theta \right)\, . 
\end{equation}
Adding these two contributions and integrating over the worldsheet,  we arrive at the Galilean superstring action
\begin{equation}\label{GSSaction}
S= -T\int\! d\tau\! \!\oint \! d\sigma \, \left\{ \sqrt{ |\dot t \bfpi_\sigma - t' \bfpi_\tau |^2}
+ \left(\dot t \bar\theta\Gamma_0 \theta' - t' \bar\theta\Gamma_0 \dot\theta \right)\right\}\, . 
\end{equation}
It is a consequence of this construction that $\theta$ is still, formally at least, a spinor of the Lorentz group, even though it transforms only under the rotation subgroup in 
the Galilean limit.  In this notation, the Galilean supersymmetry transformations are
\begin{equation}
\delta_\epsilon \theta = \epsilon\, , \qquad \delta_\epsilon {\bf x} = -i\bar\epsilon \bfG \theta\, , \qquad \delta_\epsilon t=0\, . 
\end{equation}
These transformations are the $c\to\infty$ limit of the standard relativistic supersymmetry transformations if this limit is taken for fixed
parameter $\epsilon$, and they make it clear that the Galilean supersymmetry algebra is one for which the commutator of two supersymmetry transformations is a 
space translation, rather than a spacetime translation. 

As an aside, we mention here that any attempt to scale $\theta$ in the limit that $c\to\infty$ leads either back to the ``bosonic'' Galilean string or introduces terms that blow up 
as $c\to\infty$, without any obvious means of removing them. Additionally, the option of scaling different components of $\theta$ differently (which was crucial to the results of 
\cite{Gomis:2004pw}) does not arise here because this would necessarily break the rotational invariance that is otherwise preserved by the $c\to\infty$ limit. This fact is also 
relevant to the Hamiltonian formulation of the Galilean superstring, so we leave further comment on it until  we arrive at that topic.

\subsection{Hamiltonian formulation and $\kappa$-symmetry}

The  phase-space  action of the closed Galilean superstring is
\begin{equation}\label{psact2}
\! S=  \int\!\! d\tau\! \!\oint \! d\sigma \left\{  \bfpi_\tau  \cdot{\bf p} - \dot t \varepsilon  +iTt' \bar\theta \Gamma_0\dot\theta 
- \mu \left[\bfpi_\sigma \cdot {\bf p} -t' E\right] - \frac{1}{2} \lambda \left[|{\bf p}|^2 -(Tt')^2\right]\right\}\, , 
\end{equation}
where $E$ is now related to the variable $\varepsilon$ canonically conjugate to $t$ by 
\begin{equation}
E= \varepsilon -  iT \bar\theta\Gamma_0 \theta'\, . 
\end{equation}
The equations of motion are 
\begin{eqnarray}
Dt &=&0 \, , \qquad D{\bf x} + i \bar\theta \bfG D\theta = \lambda{\bf p}\, , \qquad  \Delta  D\theta =0\, , \nonumber \\
\dot {\bf p} &=& (\mu{\bf p})'\, , \qquad \dot \varepsilon = (\mu\varepsilon +iT \bar\theta \Gamma_0 D\theta)'\, , 
\end{eqnarray}
where
\begin{equation}
D= \partial_\tau - \mu\partial_\sigma\, , 
\end{equation}
and
\begin{equation}
\Delta = \bfG\cdot {\bf p}  + T t' \Gamma_0\, . 
\end{equation}

An equivalent form of the Galilean superstring action  is
\begin{equation}\label{eqact}
S= S_{bos} + \int\! d\tau\! \!\oint \! d\sigma\,  i \bar\theta \Delta D\theta\, , 
\end{equation}
where $S_{bos}$ is the ``bosonic'' action of (\ref{psact1}). Notice that 
\begin{equation}
\Delta^2 = |{\bf p}|^2 -(Tt')^2\, , 
\end{equation}
which is zero  on-shell, by the time-reparametrization constraint. Thus, $\Delta$ is not invertible on the constraint surface. This is an indication of an additional  fermionic gauge invariance for which  there is no corresponding constraint in the action (\ref{psact2}).   

This fermionic gauge invariance  is a  Galilean version of the ``kappa-symmetry'' of the GS superstring. As in that case, the parameter $\kappa$ is an anticommuting  spacetime spinor  that depends arbitrarily on the worldsheet coordinates.  The Galilean $\kappa$-symmetry transformations are
\begin{equation}
\delta_\kappa {\bf x} = i\delta_\kappa \bar\theta \bfG\theta \, , \qquad \delta_\kappa\varepsilon =  iT \left(\bar\theta\Gamma_0\delta_\kappa\theta\right)' \, , 
\end{equation}
where
\begin{equation}
\delta_\kappa \theta = \Delta \kappa\, . 
\end{equation}
These transformations imply that
\begin{equation}
\delta_\kappa\bfpi = 2i\delta_\kappa\bar\theta \bfG  d\theta\, , \qquad \delta_\kappa E= -2iT \delta_\kappa\bar\theta \Gamma_0\theta'\, , 
\end{equation}
and  the action is invariant if we take the Lagrange multipler $\lambda$ to have the transformation 
\begin{equation}
\delta_\kappa \lambda = -4i\bar\kappa D\theta\, . 
\end{equation}

A corollary of $\kappa$-symmetry is that  the putative  orthosymplectic 2-form defined by the action (\ref{psact2})  is not invertible on the constraint surface, implying that this action  is not  in Hamiltonian form, although this can be achieved by introducing a new anticommuting spinor variable canonically conjugate to 
$\theta$ (call it $\chi$), and a new spinorial constraint that allows its elimination. This ``strictly-Hamiltonian'' action is 
\begin{eqnarray}\label{strict}
S &=&  \int\! d\tau\! \!\oint \! d\sigma \left\{  \dot{\bf x} \cdot{\bf p} - \dot t \varepsilon + i\bar\chi\dot\theta \right. \nonumber \\
&& \left. - \mu \left[\bfpi_\sigma \cdot {\bf p} -t' E\right] - \frac{1}{2} \lambda \left[|{\bf p}|^2 -(Tt')^2\right] -i \bar\xi \Phi \right\}\, , 
\end{eqnarray}
where $\xi$ is a new anticommuting spinor Lagrange multiplier for the anticommuting spinor constraint function
\begin{equation}
\Phi = \chi + \Delta\theta\, . 
\end{equation}
We may now read off Poisson brackets from this action, and then use them to compute the Poison brackets of the constraints. One finds, in particular, 
that
\begin{equation}
\left\{\Phi(\sigma), \bar\Phi(\sigma')\right\}_{PB} = -2i \Delta(\sigma) \delta(\sigma-\sigma')\, . 
\end{equation}
As already mentioned, the matrix $\Delta$ is not invertible on the constraint surface. In fact, $\Delta$  has half-maximal rank and hence two zero eigenvalues. This may be verified 
directly by choosing a set of Dirac matrices (as we shall do later) but it can also be seen from the observation that $\Delta$  is formally a momentum-space Dirac operator with null  
$4$-momentum $(Tt',{\bf p})$ and hence has half-maximal rank. The spinor constraint $\Phi=0$ is therefore a mixture of two first-class constraints (which generate the $\kappa$-symmetry gauge 
invariance) and two second-class constraints. This is similar to what happens in the Hamiltonian formulation of superparticle mechanics \cite{deAzcarraga:1982dhu,Siegel:1983hh}. 

In principle, we could solve the second-class fermionic constraints to arrive at a strictly Hamiltonian form of the action with only first-class constraints, 
but this is useful only if it can be done while maintaining manifest covariance, which is possible only if the first and second class fermionic constraints can be 
covariantly separated. For the GS string it is well-known that no Lorentz covariant separation is possible. For  the non-relativistic ``field-theory'' limit, a separation  
consistent with the symmetries preserved by that limit {\it is} possible (as follows from  the result of  \cite{Gomis:2004pw} that $\kappa$-symmetry becomes ``irreducible'' in this limit).  
In contrast, the spinor constraint $\Phi=0$ of the Galilean superstring {\it cannot} be covariantly separated into two first-class and two second-class 
constraints, even though ``covariance'' now refers only to space rotations. 

The reason for this is simple; {\it the minimal Lorentz spinor $\theta$ remains 
irreducible as a representation of the rotation subgroup of the Lorentz group}. This is true in all the dimensions for which the GS superstring exists. 
For the case in hand, we have a 4-component real spinor, which is equivalent to a 2-component complex $Sl(2;\bC)$ spinor. With respect to the $SU(2)$ 
subgroup, this becomes the intrinsically complex (albeit pseudo-real) doublet of $SU(2)$.  Thus, despite the reduction in symmetry, the status of 
Galilean string $\kappa$-symmetry is very similar to that of the GS superstring, and even more similar to that of the (relativistic) massless superparticle \cite{Siegel:1983hh}.

\section{The topological central charge}\label{subsec:GS}

The Galilean transformations of the action (\ref{psact2}) are
\begin{eqnarray}
\delta t &=& a_0\, , \qquad \delta {\bf x} = {\bf a} + {\bf v} t + {\bf w} \times {\bf x}, \nonumber \\
\delta \varepsilon &=& {\bf v}\cdot {\bf p}\, , \qquad \delta{\bf p} = {\bf w}\times {\bf p} \, , \nonumber \\
\delta \theta &=& -\frac{1}{2} {\bf w} \cdot \bfG \gamma_* \theta \, , \qquad \gamma_* \equiv \Gamma_{123}
\end{eqnarray}
and we record here the corollary that 
\begin{equation}
\delta \bar\theta = \frac{1}{2} \bar\theta\,  {\bf w}\cdot \bfG \gamma_*\, , \qquad \delta \bfpi = {\bf v} dt + {\bf w}\times \bfpi\, . 
\end{equation}
The corresponding Noether charges are 
\begin{eqnarray}
H &=& \oint \! d\sigma \, \varepsilon\, , \qquad {\bf P} = \oint \! d\sigma \, {\bf p}\, \, , \qquad {\bf B} = \oint\! d\sigma \, t{\bf p}\, , \nonumber \\
{\bf J} &=& \oint \! d\sigma\, \left[ {\bf x}\times {\bf p} - \frac{i}{2} \bar\theta\Delta\bfG\gamma_* \theta\right]\, . 
\end{eqnarray}
The action (\ref{psact2}) is also invariant under the supersymmetry transfomations
\begin{equation}
\delta_\epsilon\theta = \epsilon\, , \qquad \delta_\epsilon {\bf x} = -i\bar\epsilon \bfG \theta\, , \qquad \delta_\epsilon \varepsilon =  iT\,  \bar\epsilon \Gamma_0 \theta'\, , 
\end{equation}
for anticommuting spinor parameter $\epsilon$. A corollary is that 
\begin{equation}
\delta_\epsilon \bfpi = 0 \, , \qquad  \delta_\epsilon E=0\, . 
\end{equation}
The supersymmetry Noether charge is 
\begin{equation}\label{susyN}
Q= \oint \! d\sigma \Delta \theta\, . 
\end{equation}

These are the Noether charges, but what is their Poisson bracket (PB) algebra?  For the bosonic Galilean string, the PBs of the canonical variables may be read off from the phase-space action (\ref{psact1}) and these may  be used to compute the PB algebra of the Galilean Noether charges. The non-zero PBs are summarized by the PB relations
\begin{equation}
\left\{{\bf w}\cdot {\bf J}, \tilde{\bf w}\cdot{\bf J}_{PB} \right\}_{PB} =  {\bf w}\times \tilde{\bf w} \cdot {\bf J}\, , 
\end{equation}
which confirms that ${\bf J}$ span the $SU(2)$ algebra of space rotations, 
\begin{equation} 
\left\{{\bf P}, {\bf w}\cdot{\bf J}\right\} = {\bf w}\times {\bf P}\, , \qquad  \left\{{\bf B}, {\bf w}\cdot{\bf J}\right\} = {\bf w}\times {\bf B}\, , 
\end{equation}
which confirms that ${\bf P}$ and ${\bf B}$ are $3$-vectors, and 
\begin{equation}
\left\{H, {\bf v}\cdot {\bf B}\right\}_{PB} = {\bf v} \cdot {\bf P}\, , 
\end{equation}
which confirms that ${\bf B}$ is a generator for Galilean boosts.  As stated in the introduction, the algebra is strictly the Galilei algebra, and not the Bargmann algebra,  because the PB of ${\bf P}$ with ${\bf B}$ is zero. 

Things are not so simple for the Galilean superstring because, as already mentioned, the  PB relations of the anticommuting variables cannot be read off from the action (\ref{psact1}) because this action is not strictly in canonical form, and the strictly-Hamiltonian alternative action (\ref{strict})  involves mixed first-class and second-class constraints that cannot be separated with breaking rotational invariance. We will deal with this problem shortly by gauge fixing. However, there is a quick way to determine the PB algebra of Noether charges, as we now explain.  

First, the algebra of the Galilean Noether charges is the same as it is for the bosonic Galilean string; this is because the new anticommuting variables contribute only to the rotation generator ${\bf J}$, and its PB relations are simply statements of the $SU(2)$ representations of the 
various generators. In particular, we should expect that 
\begin{equation}\label{Qalg?}
\left\{ Q, {\bf w}\cdot {\bf J}\right\}_{PB} = -\frac{1}{2} {\bf w} \cdot\bfG \gamma_* Q\, . 
\end{equation}
The only non-obvious PB relation is that of $Q$ with itself, which we now discuss. 

As we saw earlier, the commutator of two Galilean supersymmetry transformations yields a space translation. This would lead us to expect the
supersymmetry Noether charges to have a Poisson bracket relation such that 
\begin{equation}
\{ \bar\epsilon Q, \bar\eta Q \}_{PB} =  i(\bar\epsilon \bfG\eta)  \cdot {\bf P}\, , 
\end{equation}
where $\epsilon$ and $\eta$ are two constant anticommuting spinors.  This is indeed the case for the action constructed from the Galilean string action by the replacement of $d{\bf x}$ by $\bfpi$, but the inclusion of the WZ term leads to a modification of the supersymmetry algebra, just as it does for the GS 
superstring  \cite{deAzcarraga:1989mza}. This is because the WZ Lagrangian is not strictly invariant under supersymmetry:
\begin{equation}\label{varWZ}
\delta_\epsilon\left[ \oint\! d\sigma {\cal L}_{WZ}\right] =  T \partial_\tau \oint \! d\sigma \left\{ t'\bar\epsilon\Gamma_0\theta \right\}\, .
\end{equation}
The variation of the integrand on the right hand side, with respect to a second supersymmetry transformation with parameter $\eta$, yields a modification of the Noether charge PB algebra, which is now such that 
\begin{equation}\label{QQPB}
\{ \bar\epsilon Q, \bar\eta Q \}_{PB}  =  i\bar\epsilon \left[\bfG \cdot {\bf P}  + nT \Gamma_0\right]\eta\, , 
\end{equation}
where $n$ is the winding number of the string on the $t$-circle. Thus, the product $nT$ appears as a central charge in the Galilean supersymmetry algebra. 

Before turning to the issue of gauge-fixing, we remark that the Galilean supersymmetry superalgebra spanned by $\{H,{\bf P}, {\bf G}, {\bf J}; Q\}$ is 
a sub-superalgebra of the infinite dimensional subalgebra of orthosymplectic diffeomorphisms of the phase superspace. We have not made any attempt to
characterize this full symmety algebra more explicitly, e.g. along the lines of \cite{Batlle:2016iel} for the Galilean string, because it will not play a role in this 
paper.

\section{Gauge-fixing}

Taking into account the worldsheet reparametrization invariance, we have the following gauge transformations:  for the canonical variables
\begin{eqnarray}
\delta {\bf x} &=& \alpha {\bf p} + \beta{\bf x}' -i \bar\theta \bfG\Delta \kappa\, , \qquad \delta t = \beta t'\, , \qquad \delta\theta = \beta \theta' + \Delta\kappa\, , \nonumber \\
\delta {\bf p} &=& (\beta{\bf p})' \, , \qquad \delta \varepsilon = \left[ T^2\alpha t' + \beta \varepsilon +i T\bar\theta\Gamma_0 \Delta \kappa\right]' \, , 
\end{eqnarray}
and for the Lagrange multipliers 
\begin{eqnarray}
\delta \lambda &=& \dot\alpha + \mu'\alpha- \mu\alpha' + \lambda'\beta- \lambda\beta' - 4i \bar\kappa D\theta\, , \nonumber \\
\delta \mu  &=&  \dot\beta  + \mu'\beta - \mu\beta'  \, . 
\end{eqnarray}
We may fix these gauge invariances (for $n=1$) by imposing the conditions 
\begin{equation}\label{fixit}
t= \sigma\, , \qquad \ x^3= \tau\, , \qquad \Gamma_3\theta=\theta\, . 
\end{equation}

This generalizes the gauge-fixing conditions (\ref{gfix}) except that, for simplicity of presentation, we have chosen ${\bf n}=(0,0,1)$.  A gauge transformation will preserve these conditions if
\begin{equation}\label{fixed}
\beta=0\, , \qquad p_3\alpha = - i \bar\theta \Delta \kappa \, , \qquad 
(1- \Gamma_3)\Delta \kappa =0\, , 
\end{equation}
where we now have
\begin{equation}
\Delta = \bfG \cdot {\bf p} +T\Gamma_0\, . 
\end{equation}
Notice that $\Gamma_3\theta = \theta\, \Rightarrow \,  \bar\theta \Gamma_3 =-\bar\theta$, and hence 
\begin{equation}
\Gamma_3\theta = \theta \quad \Rightarrow \quad \bar\theta = \frac{1}{2}\bar\theta (1-\Gamma_3)\, , 
\end{equation}
from which it  follows  that  $\bar\theta \Delta \kappa=0$, given the restriction on $\kappa$. In addition, we can solve
the restriction on $\kappa$ for $\kappa_-$ in terms of $\kappa_+$, where $\kappa_\pm$ are the projections of $\kappa$ onto 
the $\pm$ eigenspaces of $\Gamma_3$:
\begin{equation}
\Gamma_3 \kappa_\pm = \pm \kappa_\pm \, . 
\end{equation}
The upshot is that (\ref{fixed}) is equivalent to
\begin{equation}
\beta=0\, , \qquad p_3\alpha=0 \, , \qquad  p_3 \kappa_- =  \Delta_\perp \kappa_+\, , 
\end{equation}
where 
\begin{equation}
\Delta_\perp \equiv {\bf p}_\perp\cdot \bfG_\perp + T\Gamma_0\, . 
\end{equation}
If we assume that $p_3\ne0$ then the gauge-fixing conditions (\ref{fixit})  fix all gauge invariances except those associated to $\kappa_+$, but 
we shall see later that this residual gauge invariance acts trivially. 

Notice that the gauge-fixing condition on $\theta$ can be written as 
\begin{equation}
\theta_- =0\, , 
\end{equation}
where $\theta_\pm$ are the projections of $\theta$ onto the $\pm$ eigenspaces of $\Gamma_3$. In this notation, the gauge-fixed phase-space action 
for the Galilean superstring is 
\begin{equation}\label{supergf}
S= \int d\tau \oint d\sigma \left\{ \dot {\bf x}_\perp \cdot {\bf p}_\perp + i \bar\theta_+ \Delta_\perp \dot\theta_+ -T \sqrt{1 - T^{-2}|{\bf p}_\perp|^2}\right\}\, , 
\end{equation}
which generalizes (\ref{ppsact}).  Although this action does not involve $\sigma$-derivatives of the physical phase-space variables, these do appear in 
the Noether charge $H$, which now has a contribution from the anticommuting variables:
\begin{equation}\label{forH}
H= \oint\! d\sigma \left\{ {\bf x}'_\perp \cdot {\bf p}_\perp + i \bar\theta_+ \Delta \theta'_+\right\} \, . 
\end{equation}

\subsection{Gauge-fixed supersymmetry}

To get the supersymmetry transformations of the gauge-fixed action we have to add a compensating  $\kappa$-symmetry transformation to maintain the gauge. This
compensating $\kappa$-transformation is determined by the requirement that 
\begin{eqnarray}
0= \delta_\epsilon\theta_-  &=& \epsilon_-  + \frac{1}{2}(1-\Gamma_3)\Delta \kappa \nonumber \\
&=& \epsilon_- - p_3\kappa_- + \Delta_\perp \kappa_+\, , 
\end{eqnarray}
which implies that  
\begin{equation}\label{kapep}
p_3\kappa_-(\epsilon) =  \epsilon_- + \Delta_\perp \kappa_+\, . 
\end{equation}
We also need a compensating $\alpha$-transformation to maintain the gauge $x^3=\tau$; this is determined by the requirement that 
\begin{equation}
0= i \bar\theta\Gamma_3 \epsilon -i \bar\theta\Gamma_3 \Delta\kappa(\epsilon) + \alpha(\epsilon) p_3\, , 
\end{equation}
which tells us (given $\theta=\theta_+$) that 
\begin{equation}
p_3 \alpha  = i\bar\theta \epsilon_- + ip_3\bar\theta\kappa_- -i \bar\theta\Delta_\perp \kappa_+  = 2i \bar\theta \epsilon_-\, . 
\end{equation}

In summary, the compensating gauge transformations that are required for a supersymmetry transform to preserve the gauge conditions have 
parameters\footnote{This is a special case of a general result 
for  the  compensating gauge transformation parameters required for an arbitrary linear combination of  super-Galilean transformations to preserve 
the gauge conditions; in this general case they may depend on all super-Galilean parameters.}
\begin{equation}
\alpha(\epsilon) = \frac{2}{p_3} i\bar\theta\epsilon_- \, , \qquad 
\kappa_-(\epsilon) = \frac{1}{p_3} \left(\epsilon_- + \Delta_\perp \kappa_+\right)\, . 
\end{equation}
This result suffices for the determination of the supersymmetry transformations of the gauge-fixed theory.  There is no $\alpha$-transformation of $\theta$, so its gauge-fixed supersymmetry transformation is
\begin{eqnarray}\label{deltath}
\delta_\epsilon \theta &=& \epsilon + \Delta \kappa(\epsilon) = \epsilon_+ + \epsilon_- + \Delta \kappa_+ + \Delta\kappa_-(\epsilon) \nonumber \\
&=& \epsilon_+ + \frac{1}{p_3} \Delta_\perp \epsilon_-\, . 
\end{eqnarray}
There is both an $\alpha$-transformation and a $\kappa$-transformation of ${\bf x}_\perp$, so its gauge-fixed supersymmetry transformation is 
\begin{eqnarray}
\delta_\epsilon {\bf x}_\perp &=&  i\bar\theta\bfG_\perp \left[ \epsilon_+ - \frac{1}{p_3} \Delta_\perp \epsilon_- \right] 
+ 2\frac{ {\bf p}_\perp}{p_3} i \bar\theta \epsilon_- \nonumber \\
&=& i\bar\theta\bfG_\perp \epsilon_+ - \frac{i}{p_3} \bar\theta \left[ \bfG_\perp \Delta_\perp -2 {\bf p}_\perp\right]\epsilon_-\, . 
\end{eqnarray}
Finally, we need to consider ${\bf p}_\perp$ as this is a canonical variable in the gauge-fixed action. 
As ${\bf p}_\perp$ is  initially $\epsilon$-inert, and also $\alpha$-inert and $\kappa$-inert, its gauge-fixed supersymmetry transformation is zero. 
Notice that the undetermined $\kappa_+$ parameter has dropped out of these end results for the gauge-fixed supersymetry transformations, 
which should leave invariant the gauge-fixed action (\ref{supergf}). 

Let's first consider the $\epsilon_+$ case:
\begin{equation}
\delta \theta_+= \epsilon_+\, , \qquad \delta{\bf x}_\perp = i \bar\theta_+\bfG_\perp \epsilon_+\, . 
\end{equation}
This is indeed a symmetry of the gauge-fixed action, and the corresponding Noether charge is
\begin{equation}
Q_- =  \oint \! d\sigma\, \Delta_\perp \theta_+\, . 
\end{equation}
Next is the $\epsilon_-$ case:
\begin{equation}
\delta\theta_+ =  \frac{1}{p_3} \Delta_\perp \epsilon_- \, , \qquad \delta{\bf x}_\perp =  - \frac{i}{p_3} \bar\theta \left[ \Gamma_\perp \Delta_\perp -2{\bf p}_\perp\right] \epsilon_-\, . 
\end{equation}
To verifify that this is also a symmetry of the gauge-fixed action requires a longer calculation but the  result is that it {\it is} a symmetry, with  corresponding Noether charge 
\begin{equation}
Q_+ = \oint \! d\sigma\, p_3 \theta_+\, . 
\end{equation}
Using the fact that $\theta_+ = \Gamma_3\theta_+$, we find that 
\begin{equation}
Q_+ + Q_- = \oint \! d\sigma \Delta \theta_+\, . 
\end{equation}
This is the supersymmetry Noether charge of (\ref{susyN}) that we deduced previously, but now for gauge choice (\ref{fixit}).  Of course, this had to be so
because Noether charges are gauge invariant.  Notice that (\ref{QQPB}) can be rewritten (for $n=1$) as 
\begin{equation}
\{ \bar\epsilon Q, \bar\eta Q \}_{PB}  =  i\bar\epsilon\left[\oint\! d\sigma  \Delta (\sigma) \right]\eta\, . 
\end{equation}
An immediate corollary is that  solutions of the Galilean superstring equations of motion  for which the equation $\Delta(\sigma)\epsilon =0$ admits non-zero solutions for constant $\epsilon$ are invariant under some fraction (in fact half) of the four supersymmetries generated by the four-component  spinor charge $Q$.
We shall confirm these conclusions shortly.

\subsection{The unitarity bound}

 At this point it is convenient to choose the 4D  Dirac matrices to be of the form
\begin{equation}\label{gam3}
\Gamma_0 = \left(\begin{array}{cc} 0 & \gamma_0\\ \gamma_0& 0\end{array} \right) \, , \quad
 \bfG_\perp = \left(\begin{array}{cc} 0 & \bfg \\ \bfg & 0\end{array} \right)\, , \quad \Gamma_3 =  \left(\begin{array}{cc} \bI_2  & 0  \\ 0 & -\bI_2 \end{array}\right) \, , 
\end{equation}
where $(\gamma_0, \bfg)$ are a set of real 3D Dirac matrices, which we may assume to be such that $\gamma^{012}=1$.  For this choice we have 
\begin{equation}
\theta_+ = \left(\begin{array}{c} \vartheta \\ 0 \end{array}\right)\, , 
\end{equation}
where $\vartheta$ is a two-component spinor, nominally of  the 3D Lorentz group $Sl(2;\bR)$ although this has been broken to $SO(2)$ by the Galilean limit. 
In this notation, the action (\ref{supergf}) becomes
\begin{equation}\label{supergf2}
S= \int d\tau \oint d\sigma \left\{ \dot {\bf x}_\perp \cdot {\bf p}_\perp + i \bar\vartheta \delta_\perp \dot\vartheta  -T \sqrt{1 - T^{-2}|{\bf p}_\perp|^2}\right\}\, , 
\end{equation}
where $\bar\vartheta = \vartheta^T\gamma^0$, and the expression (\ref{forH}) for $H$ becomes 
\begin{equation}
H= \oint\! d\sigma\left\{ {\bf x}'_\perp \cdot {\bf p}_\perp + i \bar\vartheta \delta_\perp \theta'\right\}\, . 
\end{equation}

The matrix $\gamma^0\delta_\perp$ is invertible and hence we may now read off the canonical Poisson brackets from the action (\ref{supergf2}). 
Let $x^i$ and $p_i$ be the components of ${\bf x}_\perp$ and ${\bf p}_\perp$, respectively, and let $\{\vartheta^\alpha; \alpha=1,2\}$ be the 
components of $\vartheta$. Then the non-zero canonical PB relations  are
\begin{eqnarray}
\left\{x^i(\sigma), p_j(\sigma')\right\}_{PB} &=& \delta^i_j\,  \delta(\sigma-\sigma') \, , \nonumber \\
\left\{\vartheta^\alpha(\sigma),\vartheta^\beta(\sigma')\right\}_{PB} &=&   -i\frac{(\delta_\perp\gamma^0)^{\alpha\beta}}{T^2- |{\bf p}_\perp|^2} \, \delta(\sigma-\sigma')\, . 
\end{eqnarray}

We are now in a position to compute the PB relations of the supersymmetry charges. First, we observe that 
\begin{equation}
Q = \left(\begin{array}{c} q_+ \\ q_- \end{array}\right)\, , 
\end{equation}
where the two-component supersymmetry charges are 
\begin{equation}
q_+ = \oint\! d\sigma\,  p_3\vartheta\, , \qquad q_- = \oint\! d\sigma\,  \delta_\perp\vartheta\, . 
\end{equation}
Using the canonical PB relations we find that 
\begin{eqnarray}
\left\{q_+^\alpha,q_-^\beta\right\}_{PB} &=& -i\left(\gamma^0\right)^{\alpha\beta}P_3  \nonumber \\
\left\{q_\pm^\alpha, q_\pm^\beta\right\}_{PB} &=& -i \left(\bfg\gamma^0\right)^{\alpha\beta} {\bf P}_\perp - i \delta^{\alpha\beta} nT\, , 
\end{eqnarray}
where $n$ is the string winding number.  For $n=1$ this is equivalent  to the PB relation 
\begin{equation}
\left\{ Q, Q\right\}_{PB} = -i \oint\! d\sigma \Delta(\sigma) \Gamma^0\, , 
\end{equation}
which can be written in the form (\ref{QQPB}) that we deduced previously. 
 
Using the Poisson bracket to (anti)commutator prescription to pass over to the quantum theory,  and choosing a representation of the canonical commutation 
relations for which $\hat{\bf p}$ is diagonal with eigenvalues ${\bf p}$,  we arrive at the following anticommutation relation for quantum operator $\hat Q$ replacing 
the spinorial Noether charge $Q$: 
\begin{equation}
\left\{ \hat Q, \hat Q\right\} = \oint\! d\sigma \Delta(\sigma) \Gamma^0 =  \left(\bfG\Gamma^0\right)\cdot {\bf P}  +T\, \bI_4\, . 
\end{equation}
Assuming that $\hat Q$ is hermitian, and that the Hilbert space has no states of negative norm, we deduce that the matrix 
$(\bfG\Gamma^0)\cdot P  +T\, \bI_4$ is non-negative. It is easily seen (e.g. by choosing a direction for ${\bf P}$) that this implies the
(unitarity) bound 
\begin{equation}\label{bound}
|{\bf P}|^2 \le T^2\, . 
\end{equation}
This is precisely the classical bound (\ref{bogbound}) that we deduced in the introduction from the time-reparametrization constraint!

\section{Supersymmetry preservation}

The general infinitesimal transformation of $\theta$ that is gauge-equivalent to a supersymmetry  transformation is $\delta\theta = \epsilon + \beta\theta' +  \Delta\kappa$.  A bosonic configuration of the Galilean superstring; i.e. one with $\theta\equiv0$,  is supersymmetric if the equation $\delta\theta \equiv 0$ allows a non-zero solution for $\epsilon$ when $\theta\equiv 0$;  i.e. if the equation 
\begin{equation}
\epsilon + \Delta\kappa =0
\end{equation}
has a non-zero solution for (constant and uniform) $\epsilon$. If we restrict the configurations to those that satisfy the phase-space constraints then $\Delta^2=0$ and hence
\begin{equation}
\epsilon + \Delta\kappa =0 \quad \Rightarrow \ \Delta \epsilon =0\, . 
\end{equation}
This is the condition just deduced from the Galilean supersymmetry algebra.  Since $\Delta^2=0$,  a solution for $\epsilon$ will exist at any given point on the string worldsheet, but as we require the BPS condition to hold  at {\it all} points on the worldsheet,  there will not generically be a solution for constant uniform $\epsilon$. 

Given the gauge choice (\ref{fixit}), the condition $\Delta\epsilon=0$ becomes 
\begin{equation}\label{susypres}
\Delta_\perp \epsilon_\pm  = \pm p_3 \epsilon_\mp\, , 
\end{equation}
for either sign since the equation for one sign choice implies the equation for the other choice.  The lower sign equation is precisely the condition for the vanishing of the 
expression (\ref{deltath}) for the  gauge-fixed variation of $\theta$.  

For the choice of Dirac matrices in  (\ref{gam3}), we have
\begin{equation}
\epsilon_+ = \left(\begin{array}{c} \varepsilon_+ \\ 0 \end{array}\right)\, , \qquad \epsilon_- = \left(\begin{array}{c}  0\\ \varepsilon_-\end{array}\right)\, , 
\end{equation}
where  $\varepsilon_\pm$ are two-component spinors. We also have
\begin{equation}
\Delta_\perp = \left(\begin{array}{cc} 0& \delta_\perp \\ \delta_\perp &0 \end{array}\right)\, , \quad \delta_\perp = \bfg\cdot{\bf p}_\perp + T\gamma_0\, .  
\end{equation} 
Equation (\ref{susypres}) now reduces to $\delta_\perp \varepsilon_\pm = \mp p_3 \varepsilon_\mp$, which tells us that 
\begin{equation}
\epsilon =  \left(\begin{array}{c} \varepsilon_+ \\  p_3^{-1} \delta_\perp \varepsilon_+ \end{array}\right)\, . 
\end{equation}
This is a constant uniform spinor iff ${\bf p}$ is constant and uniform, which is a solution of the equations of motion. We conclude that a supersymmetric bosonic solution of the equations of motion is a string for which, in the gauge (\ref{fixit}),  the $3$-momentum density is both constant and uniform across the string worldsheet.  

Although ${\bf p}_\perp$ is constant, i.e. time-independent,  for {\it any}  solution of the equations of motion, it need not be uniform; both its direction and magnitude may vary arbitrarily with 
$\sigma$. Supersymmetric solutions are therefore very special solutions, with the property that $p(\sigma)={\bf P}$. We saw in the introduction, for the gauge $t'=1$, that such solutions are precisely those that saturate the bound $|{\bf P}|^2\le T^2$. Conversely, all solutions saturating the bound have ${\bf p}={\bf P}$. We conclude, in agreement with the conclusion based on the unitarity bound,  that {\it supersymmetric solutions are those saturating the  bound $|{\bf P}|^2\le T^2$}. 

\subsection{Supersymmetric versus non-supersymmetric strings}

To better understand the nature of supersymmetric solutions of the Galilean superstring equations of motion, it is convenient to take a step back from the fully gauge-fixed action 
to one for which only the string parametrization has been fixed by imposing the gauge condition $t'=1$. As we are concerned with bosonic solutions of the equations of motion it will suffice to return to the Galilean string action action (\ref{wind1}). Although the worldsheet time parametrization is still arbitrary in this action, it has the advantage of manifest rotational invariance. 
For $n=1$, the equations of motion, and constraint, are 
\begin{equation}\label{eom}
\dot {\bf x} = \lambda {\bf  p} \, , \qquad \dot{\bf p}= {\bf 0}\, ;  \qquad |{\bf p}|^2 = T^2\, . 
\end{equation}
We may now fix the time parametrization, and time units, by choosing \cite{Batlle:2016iel}
\begin{equation}
T\lambda(\tau,\sigma) = 1\,  \qquad \rightarrow \ {\bf p} = T\dot{\bf x}\, . 
\end{equation}
For this choice,  the generic solution of the equations of motion is  a loop of string  with a shape that will evolve in time from some arbitrary initial shape.

Let us define the ``centre of mass'' by\footnote{As mentioned in the introduction, the meaning of ``energy'' is obscure in the Galilean string context, as is the meaning of ``mass''.}
\begin{equation}
{\bf X} = \oint \! d\sigma \, {\bf x}\, . 
\end{equation}
Integration of the equations (\ref{eom}) yields the seemingly conventional point particle equations
\begin{equation}\label{cofm}
\dot {\bf P} ={\bf 0}\, ,  \qquad {\bf P}=T\dot{\bf X} \, . 
\end{equation}
As we saw in the introduction, integration of the constraint yields the bound $|{\bf P}|^2\le T^2$, and this fact is one indication that 
the centre of mass equations are far from conventional because the bound must be Galilean invariant. Indeed it is,  because the magnitude of the  3-momentum ${\bf P}$ is 
Galilean invariant, in fact super-Galilean invariant. In particular, it is unchanged  by the action of the Galilean boost generated by ${\bf B}$ because
\begin{equation}
\left\{ {\bf X}, {\bf a}\cdot {\bf P} + {\bf v}\cdot{\bf B}\right\}_{PB} = {\bf a} + \tfrac{1}{2} {\bf v}\, , 
\end{equation}
which shows that a boost merely shifts the centre of mass without changing its momentum. 

Supersymmetric strings saturate the bound, and for a parametrization such that $t'=1$ they have the property that ${\bf p}(\sigma)={\bf P}$ with $|{\bf P}|=T$. 
For a time parametrization for which $\lambda = T^{-1}$,  every point on such a string string has the  same unit velocity, $|\dot{\bf x}|=1$, so  {\it the shape of the string 
loop is time-independent} in this parametrization of the worldsheet.  This is a special feature of Galilean string solutions that are supersymmetric when considered as solutions 
of the equations of motion of the Galilean superstring\footnote{It is also reminiscent of  a ``supercurve'', which is  essentially a string with a transverse wave of arbitrary 
profile  but which can also be viewed as a string of arbitrary shape for which all points move in the same direction at the speed of light, although the transverse 
velocity is subluminal \cite{Mateos:2002yf}.}. 
Conversely,  all Galilean string solutions with this property are supersymmetric.

\section{Discussion}

There are essentially two  ``non-relativistic'' limits of the Nambu-Goto string.  One limit is a natural generalization of the standard 
low-velocity limit of a relativistic particle in which low-velocity applies to directions transverse to the string, while fluctuations of the string
continue to propagate along it  at the speed of light. This ``field-theory'' limit has been has been investigated at various times in the past two decades. 

The other limit has only recently been investigated \cite{Batlle:2016iel}.  It has some very unusual features, which we hope to have
clarified here. The limit is one in which $c$, the speed of light,  is taken to infinity for fixed  potential energy density of the string. Whereas this 
fixed quantity equals the string tension $T$ prior to the limit, the resulting Galilean string has potential energy density $T$ but  {\it zero} tension. 

Our main aim has been to apply this new limit to the Green-Schwarz superstring and thereby find an action for a 
{\it Galilean superstring}.  This turns out to be a rather simple extension of the Galilean string. Like the GS superstring, it consists of two terms, 
one of which is a Wess-Zumino term that leads to a topological charge in the  Galilean  supersymmetry algebra, and a corresponding unitarity bound 
on the energy density. However,  in contrast to the GS case, this topological charge  is a {\it central} charge in the Galilean supersymmetry algebra, and 
the unitarity bound is an {\it upper} bound on the momentum rather than a lower bound on the energy.  

In fact, this bound coincides with one that can be deduced directly from  the ``bosonic'' Galilean string action. The new feature is that bosonic solutions 
of the Galilean superstring equations of motion that saturate the bound are supersymmetric, in the sense that they are invariant under two of the four 
independent supersymmetry  transformations associated to the four-component spinor Noether charge corresponding to supersymmetry invariance of the action. 
We have shown that any supersymmetric string is a loop of string of fixed shape (in a natural parametrization) for which every point moves with constant velocity
in the same direction such as to maximise the momentum. 

The fact that supersymmetric Galilean superstrings saturate  a unitarity  bound implied by the Galilean supersymmetry algebra is reminiscent  of the 
analogous property of soliton solutions of  relativistic supersymmetric field theories; in that case we get a lower bound on the energy  in terms of a topological  
charge carried by the soliton that appears as a central charge in the standard supersymmetry algebra \cite{Witten:1978mh}. Moreover,  just as this 
bound can be understood classically as  a Bogomol'nyi-Prasad-Sommerfield (BPS) bound  \cite{Bog,Prasad:1975kr}, we have seen that the unitarity
bound of the Galilean supersymmetry algebra coincides with a bound derivable (albeit by a different method) from the classical phase-space constraints. 
However, there are as many differences as similarities. For instance, while it is true that the bound
$|{\bf P}|^2\le T^2$ is saturated by Galilean string solutions that maximize the ``energy'' $-{\bf n}\cdot P$ of the gauge-fixed action (or that minimize it for $T<0$)
there are solutions that saturate this bound for {\it any} value of $|{\bf n}\cdot P| \le |T|$.  As mentioned in the introduction, the notion of ``energy'' for a Galilean string 
is obscure, so one might wish to consider defining it to be $|{\bf P}|$ on the grounds that the bound $|{\bf P}|^2\le T^2$ would then be a closer cousin to the BPS bound, 
but we see no other reason to prefer this definition. 
 
The possibility of half-supersymmetric solutions of the equations of motion is closely related, as it is for the GS superstring, to 
a  fermionic gauge invariance, a ``$\kappa$-symmetry'', of the Galilean superstring action. As for the GS superstring, this gauge invariance
corresponds,  in a strictly-Hamiltonian approach, to  first-class fermionic constraints that are  ``mixed''  with an equal number of 
second-class fermionic constraints, in a way that precludes them being separated without breaking  symmetries of the action
(in contrast to the ``field-theory''  non-relativistic limit).  As is well known, this feature constitutes one of the main barriers to a 
covariant quantization of the GS superstring. The fact that it survives the Galilean limit suggests that the  Galilean superstring could 
serve as a useful simplified arena for investigations into the various proposals for covariant quantization of the GS 
superstring. 

Although these results were found by taking a limit of the Green-Schwarz superstring for the particular case of minimal supersymmetry in a 
4D spacetime, we believe that they are much more general. We could, of course, repeat all arguments with a different starting point, e.g. a 
10-dimensional spacetime. However, the restriction on the spacetime dimension arising in the construction of the GS action is a consequence of a 
particular Dirac matrix identity that we made no use of in our proof of the supersymmetry and $\kappa$-symmetry of the Galilean superstring. 
We therefore expect that the Galilean superstring can be constructed in any dimension, and probably with any number of supersymmetries, 
although we have not attempted to verify this. Even if this is true, it is unlikely that a similar statement holds for Galilean super p-branes with 
$p>1$, for a reason that we now briefly discuss. 

\subsection{Galilean supermembrane}

The Galilean limit of the Dirac-Nambu-Goto p-brane action of tension $T$ was found 
in \cite{Batlle:2016iel}; for arbitrary worldvolume coordinates $\xi^\mu = (\tau, \sigma^i)$, it can be written as 
\begin{equation}\label{Galp}
S= - T\int\! d\tau \int\!  d^p\sigma \, \sqrt{(\det h) h^{\mu\nu} \partial_\mu t\partial_\nu t}\, , \qquad h_{\mu\nu} =\partial_\mu {\bf x} \cdot\partial_\nu{\bf x}\, , 
\end{equation}
where $h^{\mu\nu}$ is the inverse of the $(p+1)\times(p+1)$ matrix $h$ with entries $h_{\mu\nu}$. 
The same action with 
\begin{equation}
h_{\mu\nu} \to \bfpi_\mu \cdot \bfpi_\nu\, , \qquad \bfpi_\mu = \partial_\mu {\bf x} + i\bar\theta \bfG\partial_\mu \theta \, , 
\end{equation}
is Galilean supersymmetry invariant,  but  it is not $\kappa$-symmetric and there will be no supersymmetry-preserving solutions.  

Let us consider further the $p=2$ case by taking the $c\to\infty$  limit of the relativistic supermembrane \cite{Bergshoeff:1987cm}, again in 
4D Minkowski space, for simplicity. This limit will yield a sum of the $p=2$ case of the action (\ref{Galp}) with 
a WZ  term constructed from the closed super-Galilean invariant  $4$-form 
\begin{equation}
\omega = dt \bfpi \cdot id\bar\theta \bfG \Gamma_0 d\theta \, . 
\end{equation}
This $4$-form is not obviously closed since $d\bfpi = i d\bar\theta\bfG d\theta$, and hence 
\begin{equation}
d\omega =  dt (d\bar\theta \bfG d\theta) \cdot (d\bar\theta \bfG \Gamma_0 d\theta)\ . 
\end{equation}
Not surprisingly, given the construction, this is zero as a consequence of the same Dirac matrix identities that allows the construction of the supermembrane. For coordinates  $X^m$ and anticommuting Majorana spinor $\theta$, these identities are equivalent to 
\begin{equation}
(d\bar\theta \Gamma^n d\theta)(d\bar\theta \Gamma_{nm} d\theta) \equiv 0\, .  
\end{equation}
The Galilean supermembrane that results from taking the $c\to\infty$ limit depends only on the time component of this Lorentz-vector  identity, and the restrictions 
it imposes may therefore be less severe than they are for the relativistic supermembrane (for which one finds that $D=4,5,7,11$).   
Nevertheless, it seems likely that there will be an upper bound both on $p$ and the space dimension. 
It would be interesting  to find the Galilean analog of the ``brane scan'' found for relativistic super p-branes in \cite{Achucarro:1987nc}. 

{\bf Note added}: The relation between the Galilean string and the massless Galilean particle, both massless Galilean systems, has recently been clarified in \cite{Batlle:2017cfa}.

\section*{Acknowledgements} 
We thank the Galileo Galilei Institute for Theoretical Physics for hospitality, and the INFN for partial support,  during the initial stages of research for this paper. J
G is grateful to Steven Weinberg for discussions, hospitality and support during a visit to the Physics Dept. Theory Group of the Univ. of Texas at Austin,  where this 
paper was completed; the material in it is  based upon work supported in part  by the National Science Foundation under Grant Number PHY-1620610 and with support from 
The Robert A. Welch Foundation, Grant No. F-0014. J.G also has been supported  by FPA2013-46570-C2-1-P, 2014-SGR-104 (Generalitat de Catalunya) and Consolider CPAN and by
the Spanish goverment (MINECO/FEDER) under project MDM-2014-0369 of ICCUB (Unidad de Excelencia Mar\'\i a de Maeztu).
PKT acknowledges support from the UK Science and Technology Facilities Council (grant ST/L000385/1).

\providecommand{\href}[2]{#2}\begingroup\raggedright\endgroup

\end{document}